\documentstyle[psfig]{mn}
\def\degr{\hbox{$^\circ$}}
\def\>{$>$}
\def\<{$<$}
\def\simlt{\lower.5ex\hbox{$\; \buildrel < \over \sim \;$}}
\def\simgt{\lower.5ex\hbox{$\; \buildrel > \over \sim \;$}}

\newif\ifAMStwofonts

\ifoldfss
  \ifCUPmtlplainloaded \else
    \NewTextAlphabet{textbfit} {cmbxti10} {}
    \NewTextAlphabet{textbfss} {cmssbx10} {}
    \NewMathAlphabet{mathbfit} {cmbxti10} {} 
    \NewMathAlphabet{mathbfss} {cmssbx10} {} 
  \fi
  \ifAMStwofonts
    \ifCUPmtlplainloaded \else
      \NewSymbolFont{upmath} {eurm10}
      \NewSymbolFont{AMSa} {msam10}
      \NewMathSymbol{\upi}     {0}{upmath}{19}
      \NewMathSymbol{\umu}     {0}{upmath}{16}
      \NewMathSymbol{\upartial}{0}{upmath}{40}
      \NewMathSymbol{\leqslant}{3}{AMSa}{36}
      \NewMathSymbol{\geqslant}{3}{AMSa}{3E}

    \fi
  \fi
\fi 

\ifnfssone
  \newmathalphabet{\mathit}
  \addtoversion{normal}{\mathit}{cmr}{m}{it}
  \addtoversion{bold}{\mathit}{cmr}{bx}{it}
  \newmathalphabet{\mathbfit} 
  \addtoversion{normal}{\mathbfit}{cmr}{bx}{it}
  \addtoversion{bold}{\mathbfit}{cmr}{bx}{it}
  \newmathalphabet{\mathbfss} 
  \addtoversion{normal}{\mathbfss}{cmss}{bx}{n}
  \addtoversion{bold}{\mathbfss}{cmss}{bx}{n}
  \ifAMStwofonts
    \ifCUPmtlplainloaded \else
      \UseAMStwoboldmath
      \makeatletter
      \new@mathgroup\upmath@group
      \define@mathgroup\mv@normal\upmath@group{eur}{m}{n}
      \define@mathgroup\mv@bold\upmath@group{eur}{b}{n}
      \edef\UPM{\hexnumber\upmath@group}
      \new@mathgroup\amsa@group
      \define@mathgroup\mv@normal\amsa@group{msa}{m}{n}
      \define@mathgroup\mv@bold\amsa@group{msa}{m}{n}
      \edef\AMSa{\hexnumber\amsa@group}
      \makeatother
      \mathchardef\upi="0\UPM19
      \mathchardef\umu="0\UPM16
      \mathchardef\upartial="0\UPM40
      \mathchardef\leqslant="3\AMSa36
      \mathchardef\geqslant="3\AMSa3E
    \fi
  \fi
\fi 

\ifnfsstwo
  \DeclareMathAlphabet{\mathbfit}{OT1}{cmr}{bx}{it}
  \SetMathAlphabet\mathbfit{bold}{OT1}{cmr}{bx}{it}
  \DeclareMathAlphabet{\mathbfss}{OT1}{cmss}{bx}{n}
  \SetMathAlphabet\mathbfss{bold}{OT1}{cmss}{bx}{n}
  \ifAMStwofonts
    \ifCUPmtlplainloaded \else
      \DeclareSymbolFont{UPM}{U}{eur}{m}{n}
      \SetSymbolFont{UPM}{bold}{U}{eur}{b}{n}
      \DeclareSymbolFont{AMSa}{U}{msa}{m}{n}
      \DeclareMathSymbol{\upi}{0}{UPM}{"19}
      \DeclareMathSymbol{\umu}{0}{UPM}{"16}
      \DeclareMathSymbol{\upartial}{0}{UPM}{"40}
      \DeclareMathSymbol{\leqslant}{3}{AMSa}{"36}
      \DeclareMathSymbol{\geqslant}{3}{AMSa}{"3E}
    \fi
  \fi
\fi 

\ifCUPmtlplainloaded \else
  \ifAMStwofonts \else 
    \def\upi{\pi}
    \def\umu{\mu}
    \def\upartial{\partial}
  \fi
\fi

\title[Changes in the structure of the accretion disc of EX Dra]
	{Changes in the structure of the accretion disc of \\
	 EX Draconis through the outburst cycle}
\author[R. Baptista \& M. S. Catal\'an]
       {Raymundo Baptista$^1$ and M. S. Catal\'an$^2$ \\
       $^1$ Departamento de F\'\i sica, Universidade Federal de Santa
       Catarina, Campus Trindade, 88040-900, Florian\'opolis - SC, Brazil,\\
       ~ email: bap@fsc.ufsc.br \\
 $^2$ Department of Physics, Keele University, Keele, Staffordshire,
       ST5 5BG, UK, email: msc@astro.keele.ac.uk }

\date{Accepted ????. Received ????; in original form 2000 August 31}
\pagerange{1-14}
\pubyear{2000}

\begin{document}

\maketitle

\begin{abstract}
We report on the analysis of high-speed photometry of the dwarf nova 
EX Dra through its outburst cycle with eclipse mapping techniques.
The eclipse maps show evidence of the formation of a one-armed spiral 
structure in the disc at the early stages of the outburst and reveal how 
the disc expands during the rise until it fills most of the primary Roche 
lobe at maximum light. During the decline phase the disc becomes 
progressively fainter until only a small bright region around the white 
dwarf is left at minimum light.
The eclipse maps also suggest the presence of an inward and an 
outward-moving heating wave during the rise and an inward-moving cooling
wave in the decline. The inferred speed of the outward-moving heating 
wave is of the order of 1 km\,s$^{-1}$, while the speed of the cooling 
wave is a fraction of that.
Our results suggest a systematic deceleration of both the heating and 
the cooling waves as they travel across the disc, in agreement with
predictions of the disc instability model.
The analysis of the brightness temperature profiles indicates that most 
of the disc appears to be in steady-state during quiescence and at 
outburst maximum, but not during the intermediate stages. 
As a general trend, the mass accretion rate in the outer regions is 
larger than in the inner disc on the rising branch, while the opposite 
holds during the decline branch.
We estimate a mass accretion rate of \.{M}=$10^{-8}\,M_\odot$\,yr$^{-1}$ 
at outburst maximum and \.{M}=$10^{-9.1}\, M_\odot\,$yr$^{-1}$ in 
quiescence.
The brightness temperature profile in quiescence also suggests that
the viscosity parameter is high at this stage, $\alpha_{cool}\simgt
0.25$, which favours the mass-transfer instability model.
The uneclipsed light has a steady component, understood in terms of
emission from the red secondary star, and a variable component that is
proportional to the out of eclipse flux and corresponds to about 3 per 
cent of the total brightness of the system. The variable component is
interpreted as arising in a disc wind.

\end{abstract}

\begin{keywords}
accretion, accretion discs -- binaries: close -- binaries: eclipsing --
novae, cataclysmic variables -- stars: individual: (EX Draconis).
\end{keywords}

\section{Introduction}

Dwarf novae are low-mass transfer binaries in which mass is fed to a non-magnetic ($B\simlt 10^{5}$ G) white dwarf (the primary) via an 
accretion disc by a Roche lobe filling companion star (the secondary). 
These binaries show recurrent outbursts of 2--5 magnitudes on timescales 
of weeks to months due to a sudden increase of the mass inflow in the
accretion disc.
Currently, there are two competing models to explain the cause of the 
sudden increase in mass accretion. In the mass transfer instability (MTI)
model, the outburst is the time dependent response of a viscous
accretion disc to a burst of matter transferred from the secondary
star. In this model the outburst is expected to begin near the outer 
edge of the disc and to propagate inward.
In the disc instability (DI) model, matter is transferred at a constant
rate to a low viscosity disc and accumulates in an annulus until a
critical configuration switches the disc to a high viscosity regime
and the gas diffuses rapidly inwards and onto the white dwarf. In this case,
the outburst can start at any position in the disc depending on the
radius at which the critical configuration first occurs (Smak 1984a, 
Warner 1995 and references therein).

Eclipsing dwarf novae provide an unparalleled opportunity to study the time 
evolution of non-stationary accretion discs with the aid of eclipse 
mapping techniques (Horne 1985).
Furthermore, they offer the best chance to address one of the 
major unsolved problems in accretion physics, namely, the nature of the
anomalous viscosity mechanism responsible for the inward spiraling of 
the disc material (Frank, King \& Raine 1992). 
Unfortunately, since there are only a few known eclipsing dwarf novae 
and outbursts are unpredictable events, detailed observations 
of eclipsing dwarf novae along their outburst cycle are rare.

EX Dra is a 5-hr period, eclipsing dwarf nova showing outbursts of moderate
amplitude ($\simeq 2$ mag) with a recurrence interval of $\simeq 20$ days 
(Baptista, Catal\'an \& Costa 2000, hereafter Paper I).
The spectroscopic studies by Billington, Marsh \& Dhillon (1996), Fiedler,
Barwig \& Mantel (1997) and Smith \& Dhillon (1998) led to a 
``spectroscopic'' model for the binary based on the measured radial 
velocities of the secondary star ($K_2\simeq 210-220$\,km\,s$^{-1}$) and 
of the emission lines ($K_1\simeq 163-176$\,km\,s$^{-1}$, associated with
the orbital motion of the primary star), and on the rotational broadening
of the secondary star, $v\sin i= 140$\,km\,s$^{-1}$.
In Paper I we presented and discussed a set of light curves of EX Dra in
quiescence and in outburst. The quiescent eclipse light curves were used
to derive the geometry of the binary, the masses and radii of the component
stars, as well as a distance estimate. The photometric and spectroscopic
models of the binary are consistent with each other within the uncertainties.

In this paper we report on the analysis of the light curves presented in 
Paper I with eclipse mapping techniques. We aim to follow 
the evolution of the structure of the accretion disc of EX~Dra through 
its outburst cycle. The eclipse maps capture ``snapshots'' of the disc
brightness distribution on the rise to maximum, during maximum light,
in the decline phase, and at the end of the eruption -- when the
system goes through a low brightness state before recovering its
quiescent brightness level.
Section\,\ref{analise} describes the data analysis. The results are
presented and discussed in section\,\ref{results}.
A summary of the main findings is given in section\,\ref{fim}.

\section{Data analysis} \label{analise}

\subsection{Light curves}

Our data are $V$ and $R$ band light curves of EX Dra obtained with the 
0.9-m James Gregory Telescope at the University of St.\,Andrews, Scotland,
during 1995-1996, covering four outbursts of the star. Details of the
observations and of the data reduction are given in Paper I. 

\begin{figure*}
\centerline{\psfig{figure=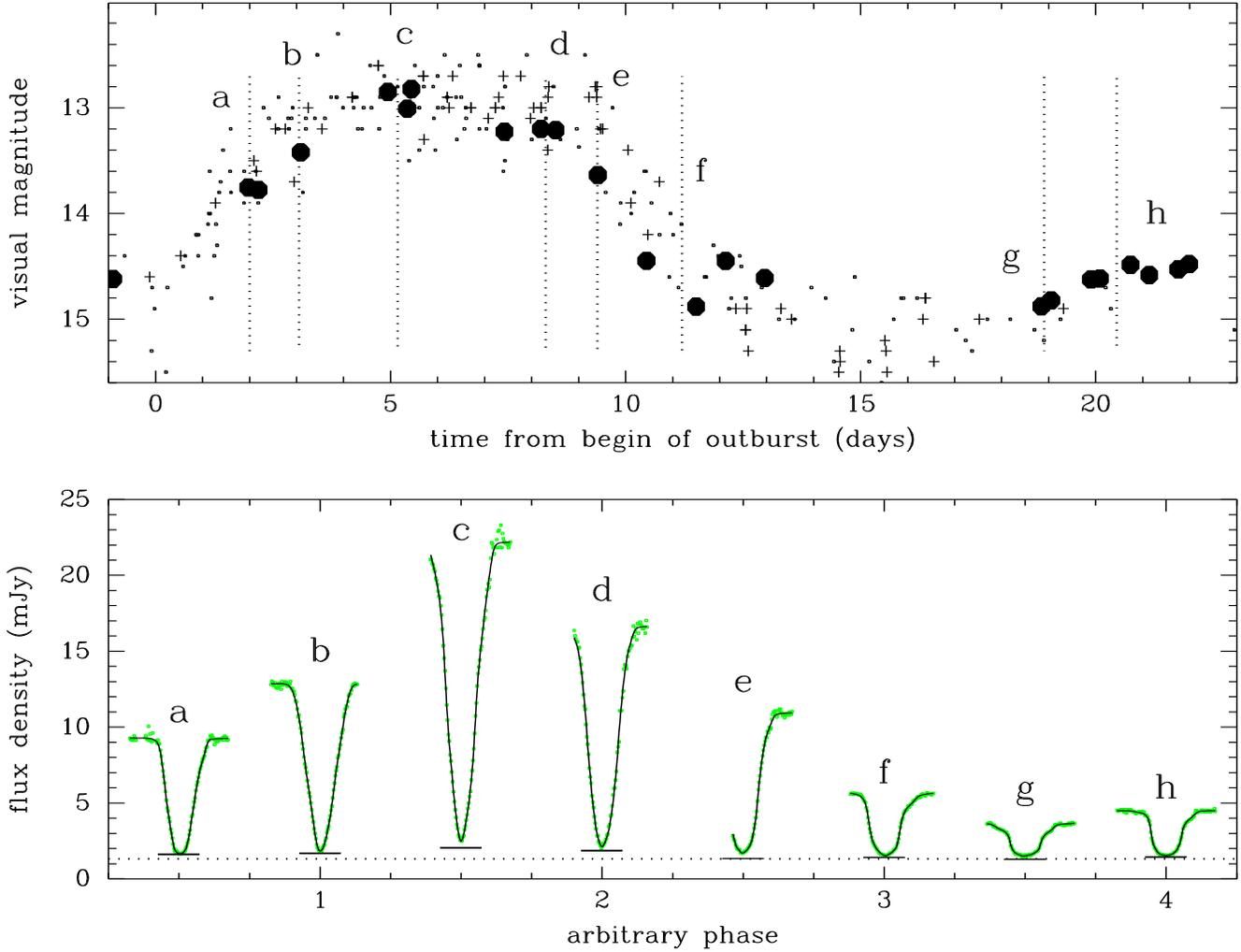,angle=-90,width=20.5cm,rheight=14.5cm}}
 \caption{ Top: superposition of visual outburst light curves of EX Dra, 
  constructed from observations made by the AAVSO and the VSNET.
  The x-axis is time relative to the onset of the outburst. Crosses
  indicate measurements of the outbursts at the epoch of our observations,
  while dots are measurements of outbursts at other epochs. Vertical dotted
  lines mark the times of the average light curves: on the rise (a-b), at
  maximum (c), during the decline (d-f), in the low state (g), and in
  quiescence (h). R-band out-of-eclipse magnitudes from our data set are
  shown as filled circles for illustration purposes. Bottom: average
  light curves seen in a sequence through the outburst cycle (dots)
  with corresponding eclipse mapping models (solid lines). The separation 
  of the light curves in the x-axis is arbitrary. Horizontal ticks show the
  uneclipsed component in each case and an horizontal dotted line marks the
  value of the uneclipsed component in the low state. }
\label{fig1}
\end{figure*}

The upper panel of Fig.\,\ref{fig1} shows a visual light curve of 
EX~Dra\footnote{constructed from observations made by the amateur 
astronomers of AAVSO and VSNET.}
obtained from the superposition of 14 outburst light curves, aligned 
according to the start of the rise to maximum.
%
Crosses indicate measurements of the outbursts at the epoch of our
observations, while dots are measurements of outbursts at other epochs. 
Only the outbursts with amplitude and duration similar to those covered by
our observations (outbursts A, B, D and E of Paper I; see fig.\,1)
were included. Filled circles mark the epochs of our observations and
indicate the corresponding $R$-band out-of-eclipse magnitudes.
These are typical type B (inside-out) outbursts, with comparable rise 
and decline timescales (Smak 1984b; Warner 1995). 
We remark that EX Dra also shows lower amplitude outbursts 
(e.g., outburst C of Paper I), and outbursts for which the rise is
significantly faster than the decline. However, these were not covered
by our observations and will not be discussed here. 
For the outbursts shown in Fig.~\ref{fig1}, 
the rise from quiescence to maximum takes about 3 days, followed by a
plateau phase of about 6 days. The decline branch lasts for 3-4 days, 
after which the star goes through a low brightness state during 4-5 days
before recovering its quiescent brightness level.

The data were grouped per outburst stage (details are given in Table
\ref{tab1}) and average light curves were obtained for eight different 
phases through the outburst cycle (marked as vertical dotted lines and
labeled from $a$ to $h$ in the upper panel of Fig.~\ref{fig1}). 
%
%
\begin{table*}
 \centering
 \begin{minipage}{90mm}
  \caption{List of average light curves.} \label{tab1}
  \begin{tabular}{@{}clccl@{}}
average & outburst & phase range & no. of & individual \\ [-0.5ex]
light curve & ~~stage & (cycle) & points & light curves \\ [0.5ex]
a & mid rise		& $-0.174,+0.174$ & 104 & 7812,7813,7817 \\
b & late rise		& $-0.174,+0.132$ & 103 & 7927 \\
c & maximum			& $-0.108,+0.174$ & 95  & 7540,7826,7827 \\
d & early decline	& $-0.099,+0.159$ & 87  & 7673,7674,9583 \\
e & mid decline 	& $-0.036,+0.174$ & 71  & 7559 \\
f & late decline	& $-0.120,+0.174$ & 99  & 7564,7970 \\
g & low state 		& $-0.135,+0.174$ & 104 & 8002,8003 \\
h & quiescence 		& $-0.174,+0.174$ & 117 & 8007,8008,8011,8013 \\
\end{tabular}
\end{minipage}
\end{table*}
For each light curve, we divided the data into phase
bins of 0.003 cycle and computed the median for each bin. The median
of the absolute deviations with respect to the median was taken as
the corresponding uncertainty for each bin. The light curves were phase
folded according to the linear plus sinusoidal ephemeris of Paper I,
\[
T_{mid} = {\rm HJD}\; 2\,448\,398.4530(\pm 1) + 0.209\,936\,98(\pm 4)\,E +
\]
\begin{equation}
+ (8.2 \pm 1.5) \times 10^{-4} \; \sin \left[ 2\pi \frac{(E-968)}{7045}
\right] \; d \;\;\; .
\label{efem}
\end{equation}

Out-of-eclipse brightness changes are not accounted for by the eclipse
mapping method, which assumes that all variations in the eclipse light curve
are due to the changing occultation of the emitting region by the
secondary star (but see Bobinger et~al. 1997 for an example of how to
include orbital modulations in the eclipse mapping scheme). Orbital
variations were therefore removed from the light curves by fitting a
spline function to the phases outside eclipse, dividing the light curve
by the fitted spline, and scaling the result to the spline function
value at phase zero. This procedure removes orbital modulations with
only minor effects on the eclipse shape itself.
The corrected average light curves are shown in the lower panel of 
Fig.\,\ref{fig1} as open squares.

\subsection{Eclipse maps} \label{mem}

The eclipse mapping method is an inversion technique that uses 
the information contained in the shape of the eclipse to recover
the surface brightness distribution of the eclipsed accretion disc.
The reader is referred to Horne (1985), Rutten et al. (1992a) and 
Baptista \& Steiner (1993) for the details of the method.

For our eclipse maps we adopted a flat grid of $75 \times 75$ pixels
centred on the primary star with side 2~R$_{\rm L1}$, where
R$_{\rm L1}$ is the distance from the  disc centre to the inner
Lagrangian point.
The eclipse geometry is determined by the mass ratio $q$ and the 
inclination $i$. We adopted the binary parameters of Paper I,
$q=0.72$ and $i=85\degr$, which correspond to an eclipse width of 
the disc centre of $\Delta\phi= 0.1085$. This combination of 
parameters ensures that the white dwarf is at the centre of the map.

The average light curves were analyzed with eclipse mapping techniques to
solve for a map of the disc brightness distribution and for the flux of an
additional uneclipsed component in each passband. The uneclipsed component
accounts for all light that is not contained in the eclipse map 
(i.e., light from the secondary star and/or a vertically
extended disc wind). The reader is referred to Rutten et~al. (1992a) and
Baptista, Steiner \& Horne (1996) for a detailed description of and 
tests with the uneclipsed component. For the reconstructions we adopted
the default of limited azimuthal smearing of Rutten et~al. (1992a), which
is better suited for recovering asymmetric structures than the original
default of full azimuthal smearing (cf. Baptista et~al. 1996).

\begin{figure*}
\centerline{\psfig{figure=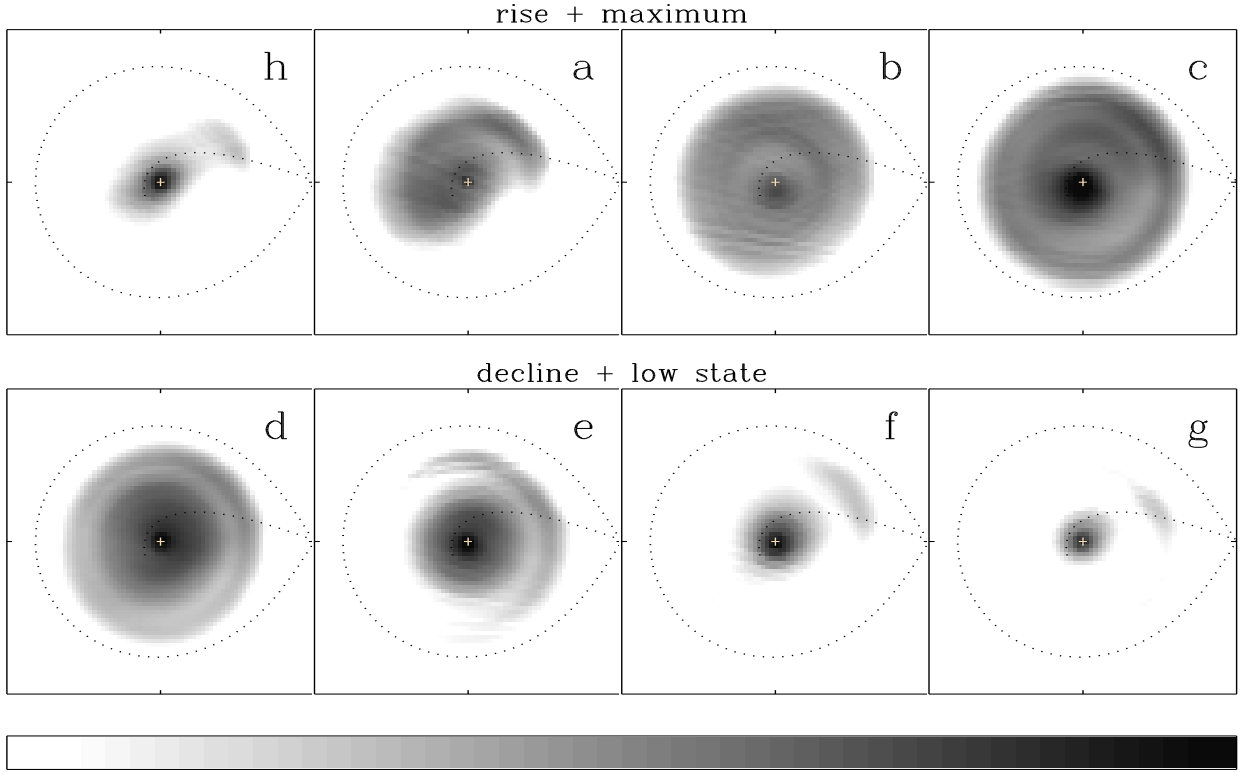,angle=-90,width=19cm,rheight=12cm}}
 \caption{ Sequence of eclipse maps of EX Dra through the outburst cycle
 in a logarithmic greyscale. Labels are the same as in Fig.\,\ref{fig1}.
 Brighter regions are indicated in black; fainter regions in white.
 A cross marks the centre of the disc; dotted lines show the Roche lobe 
 and the gas stream trajectory; the secondary is to the right of each map
 and the stars rotate counterclockwise. The greyscale bar corresponds to
 a linear scale in log of intensity from $-5.8$ to $-2.7$. }
\label{fig2}
\end{figure*}

Light curves and fitted models are shown in the lower panel of 
Fig.\,\ref{fig1}, while greyscale plots of the resulting eclipse maps 
are shown in Fig.\,\ref{fig2}. These will be discussed in detail in 
section\,\ref{results}. The reader is referred to the Appendix for 
a discussion on the reliability of eclipse mapping reconstructions for 
the case of light curves with incomplete phase coverage such as 
light curve $e$.
%
%

\section {Results} \label{results}

\subsection{Accretion disc structure} \label{structure}

A first, qualitative assessment of the evolution of the accretion disc 
of EX Dra through the outburst cycle can be made by looking at the set
of eclipse maps in Fig.~\ref{fig2}. Starting from left to right, top to
bottom, this figure shows the disc surface brightness distribution in
quiescence (h), at the early rise (a), at the late rise (b), at outburst
maximum (c), at the early decline (d), during mid-decline (e), at the late
decline (f), and in the low brightness state (g).

The sequence of eclipse maps reveal the formation of a spiral structure
at the early stages of the outburst (fig.~\ref{fig2}a), and shows that
the disc expands during the rise until it fills most of the primary Roche 
lobe at maximum light (fig.~\ref{fig2}c). 
This is in line with the observed increase in the total width of the
eclipse during the outburst (Paper~I).
The maximum of the outburst occurs when the changes in disc 
structure finally lead to a significant increase in the brightness 
of the inner disc regions (see section~\ref{radial}). 
The disc becomes progressively fainter through the decline phase,
leaving the bright spot more and more perceptible at the outer edge
of the disc (fig.~\ref{fig2}d-f). 
In the low state, the disc is reduced to a small bright region around 
the white dwarf at disc centre (possibly the boundary layer) plus a 
faint bright spot (fig.~\ref{fig2}g). 
In quiescence the disc is asymmetric, with the region along the gas 
stream trajectory being noticeably brighter than the neighbouring 
regions (fig.~\ref{fig2}h). 
The intermediate and outer regions of the accretion disc at the low state
are fainter than in quiescence suggesting that, while matter is still 
being accreted in the inner regions, mass accretion is largely reduced 
in the outer disc regions at this stage.

The detailed analysis and interpretation of the one-armed spiral
structure in the eclipse map $a$ has been presented in 
Baptista \& Catal\'an (2000).
There, it is argued that this structure is probably not tidally-induced
by the secondary star because the accretion disc radius at this stage 
is smaller than the range in radius ($0.56-0.75 \;R_{L1}$) required to
excite spiral shocks strong enough to be observed (Steeghs \& Stehle
1999). Furthermore, the tidal effect of the secondary star should, in 
principle, lead to the formation of a {\em two}-armed spiral structure 
(e.g., Yukawa, Boffin \& Matsuda 1997; Armitage \& Murray 1998)
instead of the observed one-armed spiral.
Two alternative explanations were proposed to account for the observed
spiral structure, one in terms of enhanced gas stream emission and the
other in terms of combined emission from a bright spot/stream plus the
apparently enhanced emission from the far (with respect to the secondary
star) side of a flared accretion disc. In both scenarios, the enhanced
bright spot/stream emission of map $a$ with respect to quiescence and
the smaller spiral outer radius in comparison with the quiescent disc
radius lead to the suggestion that the outbursts of EX~Dra may be driven
by episodes of enhanced mass-transfer from the secondary star. 

Joergens, Spruit \& Rutten (2000) found evidence for the presence of
spiral shocks in the accretion disc of EX Dra from Doppler tomography
close to outburst maximum. Although our maps show the disc to be asymmetric
during the mid-to-late decline, we find no evidence
of two-armed spiral structures in the accretion disc during outburst maximum
and in the decline. This could be because the spirals are relatively
weak in the continuum in comparison to the high excitation emission lines.
For example, in IP~Peg only 13 per cent of the total light in the blue
continuum is contained in the spirals against 30 per cent at C\,III+N\,III
(Baptista, Harlaftis \& Steeghs 2000).
Joergens et~al. (2000) do not quote the fraction of the total light that is
contained in the spirals in their Doppler tomograms, but mention that the
observed asymmetries are smaller than in IP~Peg, perhaps indicating that 
the shocks are weaker in EX~Dra.
Thus, it is possible that the spirals are diluted in the disc emission 
and remain undetected in the broad continuum R-band eclipse maps.

\subsection{Radial intensity distributions} \label{radial}

\begin{figure*}
\centerline{\psfig{figure=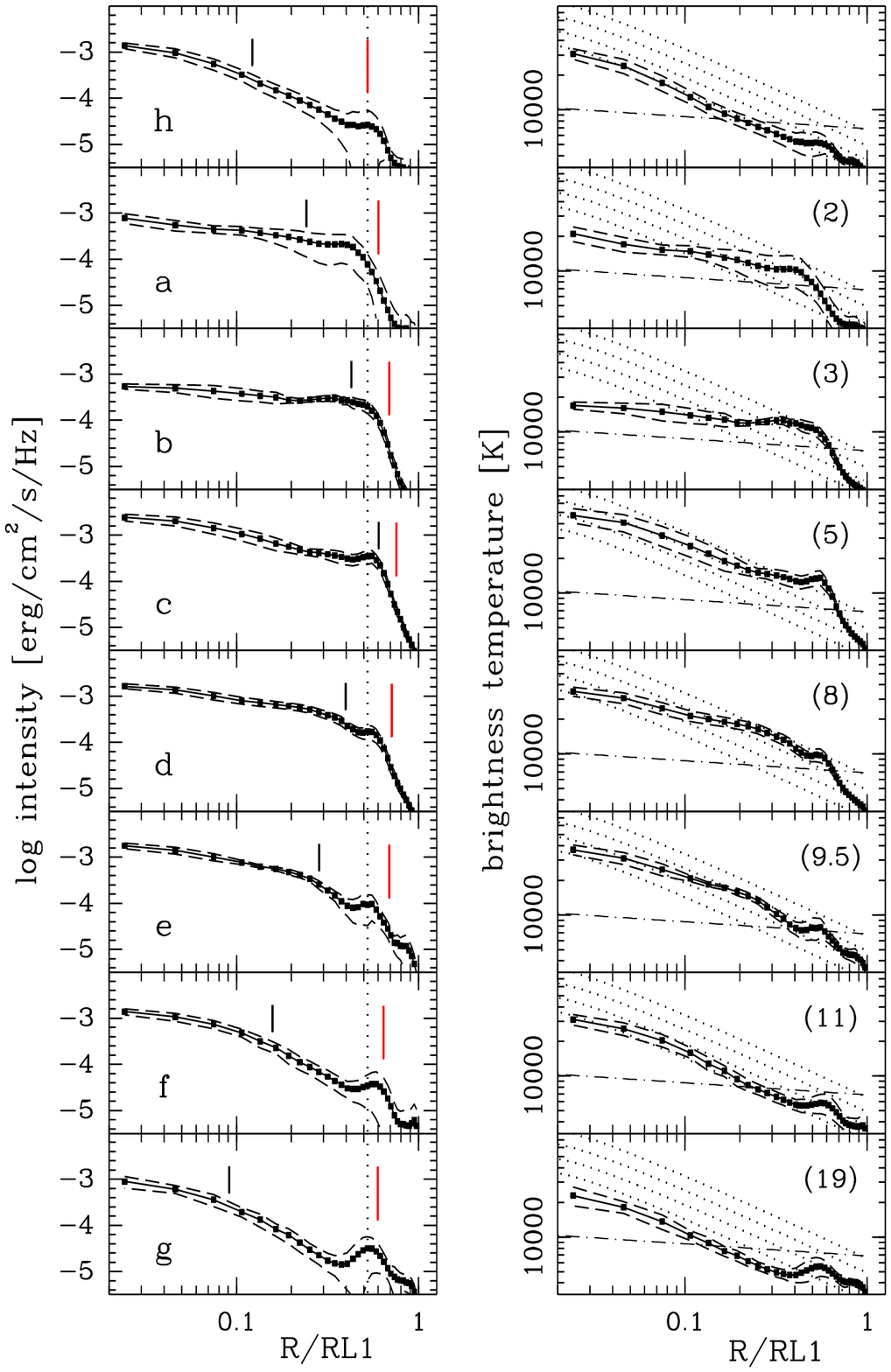,width=17cm,rheight=21cm}}
 \caption{ Left: the evolution of the radial intensity distribution through
 the outburst. Labels are the same as in Fig.\,\ref{fig1}. Dashed lines
 show the 1-$\sigma$ limit on the average intensity for a given radius.
 $R_{L1}$ is the distance from the disc centre to the inner Lagrangian point.
 A dotted vertical line indicates the radial position of the bright spot
 in quiescence. Large vertical ticks mark the position of the outer edge
 of the disc and short vertical ticks indicate the radial position at which
 the disc intensity falls below $\log I_\nu= -3.6\; (T_b\simeq 11000\;K)$.
 Right: the evolution of the radial brightness temperature distribution
 over the outburst cycle. Steady-state disc models for mass accretion rates
 of $\log$ \.{M}$= -7.5, -8.0, -8.5$, and $-9.0 \;M_\odot\;$yr$^{-1}$ are
 plotted as dotted lines for comparison. These models assume $M_1= 0.75\;
 M_\odot$ and $R_1= 0.011\;R_\odot$ (Paper I). A dot-dashed line marks the
 critical temperature above which the gas should remain in a steady,
 high mass accretion regime (Warner 1995). The numbers in parenthesis
 indicate the time (in days) from the onset of the outburst. }
\label{fig3}
\end{figure*}

A more quantitative description of the disc changes during outburst
can be obtained by analyzing the evolution of the radial intensity
distribution.
The left-hand panels of Fig.~\ref{fig3} show the evolution of the radial
intensity distribution along the outburst. We divided the eclipse
maps in radial bins of $0.03\;R_{L1}$ and computed the median
intensity at each bin. These are shown in Fig.~\ref{fig3} as 
interconnected squares. 
%
The dashed lines show the 1-$\sigma$ limits on the average intensity. 
The labels are the same as in Fig.~\ref{fig1}.
The large dispersion seen in the intermediate regions of map $a$, and in 
the outer regions of maps $f, g$ and $h$ reflect the large asymmetries 
present in these eclipse maps (e.g. spiral structure, bright spot).

In order to quantify the changes in disc size during outburst we 
arbitrarily defined the outer disc radius in each map as the radial
position at which the intensity distribution falls below 
$\log I_\nu({\rm radius}) = -4.6$, which corresponds to the maximum
intensity of the bright spot in the eclipse map in quiescence, $h$. 
The computed outer disc radius is shown as a large vertical tick in
each panel. As a reference, the radial position of the quiescent bright
spot is marked by a vertical dotted line.

During outburst the disc expands by 50 per cent, from $0.51\,R_{L1}\;
(=0.28\;a$, where $a$ is the binary separation) in quiescence to 
$0.75\,R_{L1}$ ($=0.41\,a$) at outburst maximum. 
The disc radius at outburst maximum is larger than the tidally limited 
radius for a mass ratio $q=0.72$, $R_d(max)=0.66\;R_{L1}$ (Paczy\'{n}ski 
1977), and is large enough to allow the tidal effect of the secondary 
star to induce spiral shocks in the outer disc (Steeghs \& Stehle 1999),
supporting the results of Joergens et~al. (2000).
On the other hand, if the disc expands up to the 3:1 resonance radius
(which is $r_{32}= 0.75\;R_{L1}$ for $q=0.72$, see Warner 1995),
then one would expect to start seeing superhumps. 
Unfortunately this is hard to test with our data set because most
of the outburst light curves cover a narrow phase range around the 
eclipse, although the only outburst light curve with broader out-of-eclipse
phase coverage (cycle 7540) shows evidence of a 15 per cent modulation 
reminiscent of a superhump (Paper~I, see fig.\,3).
Further outburst photometry is required in order to test this possibility.
The disc shrinks steadily along the decline and during the following 
quiescent phase. In the low state, some eight days after the end of the
outburst, the disc is still larger than in quiescence, with an outer
radius of $0.6\;R_{L1}$.

The DI model predicts that heating and cooling wave-fronts propagate 
through the disc during the transitions between the low-viscosity 
quiescent state and the high-viscosity outbursting state (e.g., Lin, 
Papaloizou \& Faulkner 1985).
Simulations aimed to test whether it is possible to detect these
transition fronts with eclipse mapping show that the radial position
of the front can be properly recovered from the eclipse map, but the
sharp break in the slope of the radial temperature distribution associated
with the transition front is blurred because of a radial smearing
effect caused by the entropy (Bobinger et~al. 1997). Hence, while
time-resolved eclipse mapping through an outburst can be used to
track the movement of transition fronts, one should not expect to
see the sharp breakes in the slope of the radial temperature
distribution associated with these fronts.

In order to test for the presence and to trace the movement of 
transition fronts in the accretion disc, we arbitrarily defined a 
reference intensity level of $\log I_\nu ({\rm front})= -3.6$,
corresponding to a brightness temperature of $T_b \simeq 11000$~K.
This lies in the range of temperatures where the disc gas is
expected to be thermally and viscously unstable in the DI model
(e.g., Warner 1995) and should yield a reasonable tracer of possible 
transition fronts.
Changing the reference intensity level by $\simeq 15$
per cent does not significantly change the results below.
The radial position at which the intensity distribution falls below the
reference intensity level $I_\nu ({\rm front})$ is marked by a small
vertical tick in the left-hand panels of Fig.~\ref{fig3}. 

It is seen that the radial position of the reference intensity changes
significantly along the outburst, moving outward on the rising branch 
and inward along the decline. These changes suggest the propagation
of an outward-moving heating wave on the rise and of an inward-moving
cooling wave on the decline. The outward-moving heating wave 
reaches the outer regions of the disc at outburst maximum 
by the time the brightness of the inner disc has increased.
If the brightening of the inner disc is interpreted as the consequence
of the arrival of an inward-moving heating wave at disc centre, this
suggests that the inward and outward moving heating waves have similar
velocities, since the involved timescales and the distances traveled by
the two waves are comparable. 
The inward-moving cooling wave seems already on its way to the disc
centre by the end of the plateau phase (eclipse map $d$).
In the low state the reference intensity level has moved closer to
disc centre than in quiescence, in agreement with the conclusions drawn
in section~\ref{structure}.

Assuming that the changes in the radial position of the reference
intensity level represent the changes in position of the heating and
cooling waves, we used the measured positions together with the inferred 
time interval between consecutive eclipse maps to estimate the velocities 
of the waves along the outburst. The results are summarized in
Table~\ref{tab2}. 
%
\begin{table*}
 \centering
 \begin{minipage}{90mm}
  \caption{Measuring the speed of the transition waves. $R_f$ is
   the radius at which the intensity falls below $\log I_\nu= -3.6 \;
   (T_b\simeq 11000\;K)$; $\Delta t$ is the time interval between the
   two consecutive maps; $v_f (= \Delta R_f/\Delta t)$ is the speed of 
   the moving wave (in $km\;s^{-1}$).} \label{tab2}
  \begin{tabular}{@{}ccccc@{}}
outburst & $R_f/R_{L1}$ & $\Delta R_f/R_{L1}$ & $\Delta t$ & $v_f$ \\
 stage &&& (days) & ($km\;s^{-1}$) \\ [0.5ex]
h $\mapsto$ a &
$0.12 \mapsto 0.24$ & $+0.12\pm 0.03$ & $2.0\pm 0.3$ & $+0.41\pm 0.12$ \\
a $\mapsto$ b & 
$0.24 \mapsto 0.43$ & $+0.19\pm 0.05$ & $1.0\pm 0.2$ & $+1.30\pm 0.43$ \\
b $\mapsto$ c & 
$0.43 \mapsto 0.60$ & $+0.17\pm 0.05$ & $2.2\pm 0.3$ & $+0.53\pm 0.17$ \\
c $\mapsto$ d & 
$0.60 \mapsto 0.40$ & $-0.20\pm 0.04$ & $3.2\pm 0.3$ & $-0.43\pm 0.10$ \\
d $\mapsto$ e & 
$0.40 \mapsto 0.28$ & $-0.12\pm 0.04$ & $1.1\pm 0.2$ & $-0.75\pm 0.26$ \\
e $\mapsto$ f & 
$0.28 \mapsto 0.16$ & $-0.12\pm 0.03$ & $1.8\pm 0.8$ & $-0.46\pm 0.23$ \\
f $\mapsto$ g &
$0.16 \mapsto 0.09$ & $-0.07\pm 0.03$ & $7.7\pm 0.9$ & $-0.06\pm 0.03$ \\
g $\mapsto$ h &
$0.09 \mapsto 0.12$ & $+0.03\pm 0.02$ & $1.5\pm 0.6$ & $+0.14\pm 0.10$ \\
\end{tabular}
\end{minipage}
\end{table*}
The uncertainties in the derived velocities are dominated by the
uncertainties in the time interval between consecutive eclipse maps.
The speed for the interval ($h \mapsto a$) was computed under
the assumption that the eclipse map in quiescence, $h$,
corresponds to the disc brightness distribution at the onset of the
outburst and, therefore, has to be interpreted with care. 
One should also be wary of the derived speed for the interval
($c \mapsto d$) as it is not possible to know for sure whether the
inward-moving cooling wave appeared at the same time (and radial
position) at which the heating wave reached the outermost regions of
the disc. Since it may well be possible that the inward-moving cooling
wave appeared only at the end of the plateau phase (e.g., Cannizzo 1993a),
the derived value of the speed in this case should be a lower limit.

The inferred speed of the heating wave is of the order of 
1\,km\,s$^{-1}$, while the speed of the cooling wave is a fraction 
of that, in reasonable agreement with the predictions from numerical
simulations of transition fronts in accretion discs (Menou et~al. 1999). 
Nevertheless, the inferred speed of the heating wave, $v_f(hot)=
0.5-1.3$\,km\,s$^{-1}$, is a factor of about two smaller than  
predicted by Menou et~al. (1999), which may suggest that, if the DI
model is right, the viscosity parameter in the low state is 
$\alpha_C\simgt 0.02$ or that the radial transport of energy 
inside the heating front is not very efficient.
The derived speed of the cooling wave, $v_f(cold)= 0.46-0.75$\,km\,s$^{-1}$, 
is in between those inferred by Bruch, Beele \& Baptista (1996) for 
OY~Car ($v_f \simeq0.14$ \, km\,s$^{-1}$) and by Bobinger et~al. (1997) 
for IP~Peg ($v_f \simeq 0.8$\,km\,s$^{-1}$).

Perhaps more important, we observe a systematic {\em deceleration} of
both the heating and the cooling waves as they travel across the disc
(at the 2-$\sigma$ level for the heating wave and at the 1-$\sigma$
level for the cooling wave), in good accordance with the predictions 
of the DI model (Menou et~al. 1999). 
This seems a subtle but important evidence in favour of the disc 
instability model, although it should be looked with certain 
reservation given the relatively low statistical significance of the
result.

In terms of the $\alpha$-disc formulation of Shakura \& Sunyaev (1973),
the non-dimensional viscosity parameter of the high state, $\alpha_{H}$, 
can be written as the ratio of the speed at which the heating front travels
across the disc, $v_f(hot)$, and the sound speed inside the heating front,
$c_s$ (Lin et~al. 1985, Cannizzo 1993b),
\begin{equation}
\alpha_{H} \approx \frac{v_f(hot)}{c_s}
 = 0.082\, \left[ \frac{v_f(hot)}{km\,s^{-1}} \right]
 \left[ \frac{T_f}{18000\;K} \right]^{-1/2} \; ,
\end{equation}
where $T_f$ is the temperature of the heating front. Assuming 
$T_f\!=\!18000\;K$ (Menou et~al. 1999), we find $\alpha_{H}\!\simeq
\!0.11\; (v_f(hot)/1.3$\,km\,s$^{-1}$), in good agreement with the 
values derived for other dwarf novae (cf. Warner 1995).

\subsection{Radial brightness temperature distributions} \label{trad}

A simple way of testing theoretical disc models is to convert the
intensities in the eclipse maps to blackbody brightness temperatures,
which can then be compared to the radial run of the effective temperature
predicted by steady state, optically thick disc models. However, as
pointed out by Baptista et~al. (1998), a relation between the effective
temperature and a monochromatic brightness temperature is non-trivial,
and can only be properly obtained by constructing self-consistent models
of the vertical structure of the disc. Unfortunately, since we have
only $R$ band light curves, a detailed disc spectrum modeling is beyond
the reach with our data set. Therefore, our analysis here has to be
considered carefully.

The right-hand panels of Fig.~\ref{fig3} show the evolution of the 
disc radial brightness temperature distribution along the outburst
in a logarithmic scale.
The blackbody brightness temperature that reproduces the observed 
surface brightness at each pixel was calculated assuming 
a distance of 290~pc to EX~Dra (Paper I). 
The eclipse maps were then divided in radial bins of $0.03\;R_{L1}$ 
and a median brightness temperature was derived for each bin. 
These are shown in Fig.~\ref{fig3} as interconnected squares. 
The dashed lines show the 1-$\sigma$ limits on the average temperatures.
Steady-state disc 
models for mass accretion rates of $10^{-7.5}$, $10^{-8}$, $10^{-8.5}$
and $10^{-9}\; M_\odot$\,yr$^{-1}$ are plotted as dotted lines for 
comparison. These models assume $M_1= 0.75 \; M_\odot$ and $R_1= 0.011
\; R_\odot$ (Paper I). The number in parenthesis indicate the time (in
days) from the onset of the outburst.

The brightness temperatures in the quiescent disc range from $5000\;K$
in the outer disc (at $R=0.5\; R_{L1}$) to about $28000\;K$ in the 
inner disc ($R=0.05\; R_{L1}$), leading to a mass accretion rate
of \.{M}= $10^{-9.1 \pm 0.3}\; M_\odot\,$yr$^{-1}$, independent of
disc radius. We note that the brightness distribution at the
inner disc in quiescence is dominated by the light from the compact
central source. The corresponding brightness temperatures are in
agreement with that inferred for this source in Paper I.
Two days after the onset of the outburst, the temperatures in the 
intermediate disc regions (the site of the spiral seen in the eclipse
map $a$) have risen from $\simeq 6000\;K$ to $\simeq 11000\;K$. 
Along the rise, the radial temperature distribution is flatter than 
the $T \propto R^{-3/4}$ law expected for steady mass accretion, 
leading to larger mass accretion rates in the outer disc than in the
inner disc regions. This is expected for both the DI and MTI models.
In the DI model, the intermediate and outer regions would have already 
switched to the high-viscosity, high mass accretion regime while the 
inner disc would still be in the low-viscosity, low mass accretion state.
In the case of the MTI model, the burst of transferred matter would be 
diffusing inward on a viscous timescale from its initial site in the
intermediate/outer disc while the mass accretion rate in the inner disc
would still be low.

At outburst maximum the temperatures range from $12000\;K$ in the outer 
disc (at $R=0.6\; R_{L1}$) to $\simeq 40000\;K$ in the inner disc regions
($R=0.05\; R_{L1}$), and are in reasonably good agreement with the 
$T \propto R^{-3/4}$ law.
The inferred mass accretion rates are \.{M}= $10^{-8.0 \pm 0.3}\;
M_\odot\,$yr$^{-1}$ at $R= 0.1\; R_{\rm L1}$ and $10^{-7.6 \pm 0.2}\; 
M_\odot\,$yr$^{-1}$ at $R= 0.3\; R_{\rm L1}$.
During the decline, the radial temperature distribution departs from the
$T \propto R^{-3/4}$ law in the direction expected for an evolving disc
(Mineshige 1991), with the temperatures progressively falling back to 
their quiescent values as the cooling wave moves inward.
As a general trend, the mass accretion rate in the outer regions 
($R=0.3\; R_{L1}$) is larger than in the inner disc ($R=0.1\; R_{L1}$)
on the rising branch, while the opposite holds during the decline branch.
The different values of $R_{L1}$ and $M_1$ in the literature 
(e.g, Billington et~al. 1996; Fiedler et~al. 1997; Smith \& Dhillon 1998)
introduce an uncertainty in the inferred mass accretion rates of
$\Delta\log$\,\.{M}$=0.04$, small in comparison with the derived 
uncertainties.
The inferred \.{M} at outburst maximum, $(10-25)\times 10^{-9}\;
M_\odot\,$yr$^{-1}$, is larger than the mass accretion rate estimates 
derived for nova-like systems of similar orbital period: $(1.6-10)\times
10^{-9}\; M_\odot\,$yr$^{-1}$ (Horne 1993; Rutten et~al. 1992a; 
Baptista et~al. 1995, 1996).

According to the DI model, 
there is a critical effective temperature, $T_{\rm eff}(crit)$,
below which the disc gas should remain while in quiescence in order 
to allow the thermal instability to set in, and above which all the 
disc gas should stay while in outburst (e.g., Warner 1995),
\begin{equation}
T_{\rm eff}(crit) = 7476\;\left( \frac{R}{R_{L1}} \right)^{-0.105}
\left( \frac{M_1}{0.75\,M_\odot} \right)^{-0.15} \; K \;\; .
\label{tcrit}
\end{equation}
This relation is plotted in the right-hand panels of Fig.~\ref{fig3}
as a dot-dashed line.
It can be seen that the disc brightness temperatures are well above 
those predicted by the relation (\ref{tcrit}) during outburst rise, 
maximum, and early decline (eclipse maps $a-d$). The temperatures in
the outer disc start to fall below $T_{\rm eff}(crit)$ at the late decline,
while most of the accretion disc ($R>0.2\;R_{L1}$) is below 
$T_{\rm eff}(crit)$ in the low state and in quiescence.
As pointed out above, the intensity distribution of the inner disc regions
($R\simlt 0.2\;R_{L1}$) in quiescence is dominated by light from the 
compact central source and, therefore, the radial temperature distribution 
of the eclipse map $h$ may not be reliable in its innermost regions.

\subsection{The quiescent disc} \label{quiescent}

In order to test if the temperatures in the quiescent disc are everywhere
below those predicted by the relation (\ref{tcrit}), we separated the 
contribution of the compact central source (CS) from the light curve $h$
with a procedure similar to that used in Paper I to measure the 
contact phases of the bright spot. 
In doing this, we are assuming that the compact central source is
a separate light source from the accretion disc itself (e.g, the
boundary layer plus the white dwarf).

The ingress/egress of the CS are seen as those intervals where 
the derivative of the light curve is significantly different from zero.
A spline function is fitted to the remaining regions in the derivative 
to remove the contribution from the extended and slowly varying eclipse 
of the disc. Estimates of the CS flux are obtained by integrating 
the spline-subtracted derivative at ingress and egress. The light curve 
of CS is then reconstructed by assuming that the flux is zero between 
ingress and egress and that it is constant outside eclipse.
The separated light curve of the accretion disc is obtained by 
subtracting the reconstructed CS light curve from the original light curve.
We performed a similar analysis on the individual light curve 8012
(Paper I) in order to separate the light curve of the quiescent disc 
in the $V$ band. The derived fluxes of the central source in the $V$
and $R$ band light curves in quiescence are, respectively, 0.7~mJy and
0.55~mJy.

\begin{figure*}
\centerline{\psfig{figure=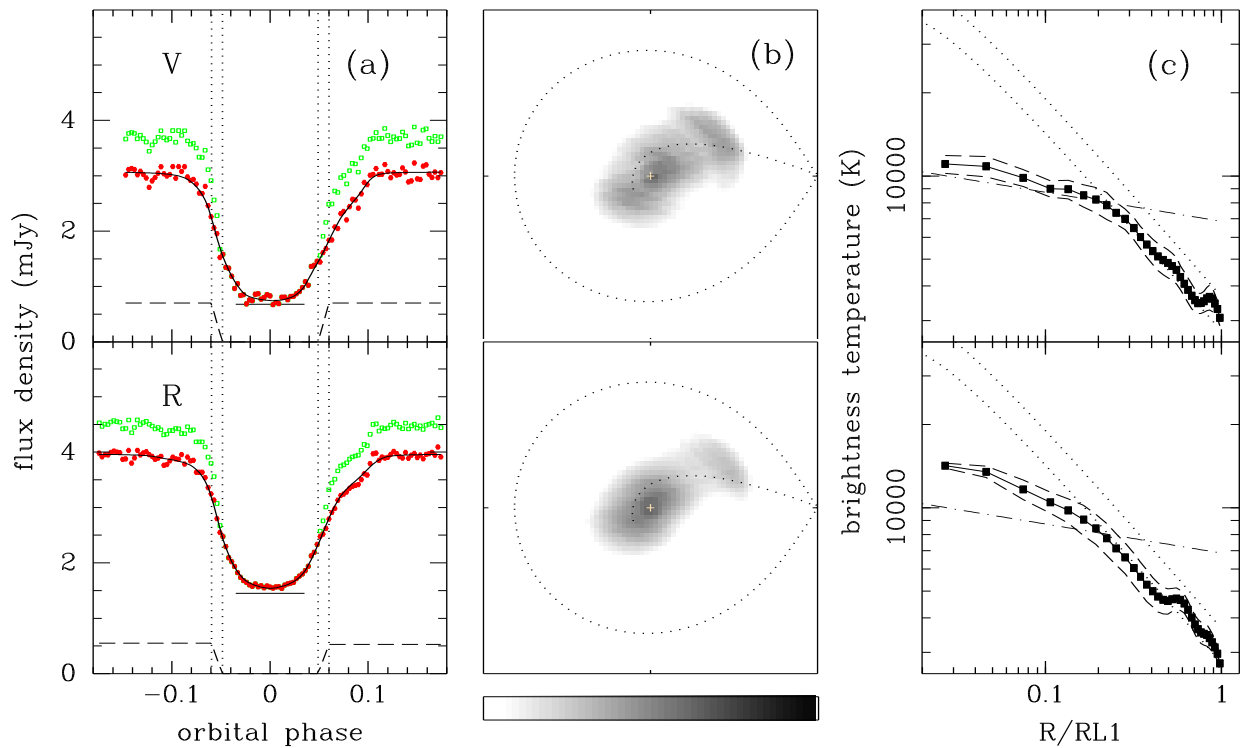,angle=-90,width=18cm,rheight=12cm}}
 \caption{ $V$ (top) and $R$ (bottom) band eclipse mapping in quiescence.
 (a) The original light curves are shown as open squares, the reconstructed
 light curves of the central source as dashed lines, and the separated disc
 light curves as filled circles. Eclipse mapping models obtained for the 
 disc light curves are shown as solid lines. Horizontal ticks mark the
 uneclipsed component in each case. Vertical dotted lines mark the ingress
 and egress phases of the central source. (b) Eclipse maps in a
 logarithmic greyscale. The notation is the same as in Fig.~\ref{fig2}.
 (c) The radial brightness temperature distributions. Steady-state 
 disc models for mass accretion rates of $\log$ \.{M}$= -8.5$, and $-9.0
 \;M_\odot\,$yr$^{-1}$ are plotted as dotted lines for comparison. These 
 models assume $M_1= 0.75\; M_\odot$ and $R_1= 0.011\;R_\odot$ (Paper I). 
 A dot-dashed line marks the critical temperature above which the gas 
 should remain in a steady, high mass accretion regime (Warner 1995). 
 The notation is the same as in Fig.~\ref{fig3}. }
\label{fig4}
\end{figure*}

The left-hand panels of Fig.~\ref{fig4} show the original light curves 
(open squares), the reconstructed light curve of the central source
(dashed lines), and the separated disc light curves (filled circles). 
Eclipse mapping models obtained for the disc light curves are shown as
solid lines. Horizontal ticks mark the uneclipsed component in each case.
The middle panels show the resulting eclipse maps in the same logarithmic
greyscale as in Fig.~\ref{fig2}, while the right-hand panels show the
corresponding radial brightness temperature distributions.
In order to obtain a clean picture of the disc temperatures in quiescence,
the temperature distributions were computed for the disc regions
excluding the quadrant that contains the bright spot and the gas stream
trajectory (the upper right quadrant in the eclipse maps of
Fig.~\ref{fig4}).
%

The $V$ band eclipse map is noisier than the $R$-band map because
the former is derived from only one light curve, while the later is the
result of the analysis from an average of four light curves.
The morphology of the two eclipse maps is the same, showing a conspicuous
bright spot at disc rim plus enhanced emission along the gas stream
trajectory. 
The radial temperature distributions yields similar brightness temperatures
in the outer and intermediate disc regions, but the $R$-band map leads
to larger temperatures in the inner disc ($R\simlt 0.2\; R_{L1}$).
We checked whether this difference could be caused by an underestimate 
of the flux of the CS in the $R$-band, leading to residual CS light in 
the inner regions of the eclipse map.
We found that even relatively small increases in the assumed flux of 
CS result in disc light curves with reversal brakes in slope during
ingress/egress, signaling that too much flux was being removed from 
the light curve.
Therefore, the differences in brightness temperatures seem real.

After the removal of the central source, the radial temperature 
distributions become flatter than the $T \propto R^{-3/4}$ law in
the inner disc regions. However, the inner disc of EX~Dra is still
hotter than the rest of the disc.
The inferred temperatures for $R<0.2\;R_{L1}$ are above 
$T_{\rm eff}(crit)$, indicating that the inner disc regions are 
stable against the thermal limit cycle of the DI model.
The brightness distributions in the outer disc regions are consistent 
with a steady-state disc at a mass accretion rate of \.{M}= $10^{-9.1
\pm 0.3}\; M_\odot\,$yr$^{-1}$, as inferred in section~\ref{trad}.
This is a lower limit to the mass transfer rate in the binary.
The fact that the inferred brightness temperatures in $V$ and $R$ are
different for $R<0.2\;R_{L1}$ may be an indication that the inner 
disc is optically thin, while the consistency between the $V$ and $R$ 
brightness temperatures for $R>0.2\;R_{L1}$ suggests that the outer 
disc is optically thick in quiescence.

On the other hand, according to the DI model, if the inner disc regions
are always hotter than $T_{\rm eff}(crit)$ they should remain in a high
viscosity steady-state. Therefore, it is possible that what was called
the compact central source is simply the innermost, brightest (and 
steady-state) regions of the quiescent accretion disc of EX Dra, 
the steep brightness distribution of which leads to the sharp changes 
in the slope of the light curve that was associated to the CS. 
In this case, there should be no subtraction of a CS component from the 
light curve and the correct radial temperature distribution in quiescence
is that shown in Fig.~\ref{fig3}.
This is in line with the observed variability of CS both in size and in 
brightness and with the fact that the inferred size of CS is significantly
larger than the expected size of the white dwarf in EX~Dra (Paper~I).

If this scenario is true, EX~Dra is the first eclipse mapped dwarf nova 
to show a steady-state disc in quiescence (the other, shorter period 
dwarf novae show a flat radial temperature distribution in quiescence 
which is clearly not consistent with a steady-state disc, e.g., Wood 
et~al. 1986, 1989; Wood, Horne \& Vennes 1992). 

This scenario is supported by numerical simulations of accretion discs 
under the DI framework, which indicate that if the mass transfer rate is
relatively high, \.{M}$_2 \simgt 10^{-9} \;M_\odot$\,yr$^{-1}$, the 
inner disc regions are sufficiently hot to permanently stay in a high
viscosity, optically thick steady-state (Papaloizou, Faulkner \& Lin 1983; 
Lin et~al. 1985; Truss et~al. 2000). 
These simulations also predict that, in these cases, the outburst starts 
at the intermediate disc regions, in the interface between the inner hot 
and the outer cool disc -- in agreement with our observations (see
section~\ref{radial}).

However, the assumption that the correct radial temperature distribution
in quiescence is that shown in Fig.~\ref{fig3} leads to another 
interesting line of reasoning. The close match of the observed temperature
distribution to the $T \propto R^{-3/4}$ law not only in the inner disc
regions but at all radii suggests that the quiescent disc is in a 
steady-state everywhere (the observed flattening of the distribution
at large disc radii, $R\simeq 0.5\;R_{L1}$, is probably caused by the
bright spot). Taking the time interval from the end of the outburst to
quiescence (map $h$) as an upper limit to the viscous timescale for
the disc to reach the steady-state ($t_\nu \simlt 10$ days), the 
viscosity parameter in quiescence, $\alpha_C$, can be estimated (Warner 
1995),
\begin{equation}
\alpha_C= \left( \frac{r}{H} \right)^2 \frac{1}{t_\nu \, \Omega_k(r_d)}
\simgt 0.25 \;\;\; ,
\end{equation}
where $\Omega_k(r_d)= 1.9\times 10^{-3}\; s^{-1}$ is the Keplerian
frequency at the disc radius ($r_d = 0.43\; R_\odot$, see Paper~I),
$H$ is the disc scale height, and we adopted $(r/H)=0.05$ for an
\.{M}$=10^{-9.1}\; M_\odot\; yr^{-1}$ (Mayer \& Mayer-Hofmeister 1982; 
Smak 1992).

This result is in clear disagreement with the DI model, which requires
that the viscosity parameter in quiescence be significantly lower
than that during outburst, $\alpha_C \simeq 0.25\;\alpha_H \simlt 0.03$
(in order to achieve realistic outburst model light curves, e.g.,
Smak 1984c; Mineshige 1988) and predicts that the discs of quiescent
dwarf novae should hardly reach a steady-state because the (viscous)
timescale required to achieve a stationary configuration with such a
low $\alpha$ ($\simgt 10^2$ days) is generally much longer than the
time interval between outbursts (e.g., Warner 1995).

Thus, if the radial temperature distribution of map $h$ is correct,
the resulting high-viscosity, steady-state quiescent disc of EX~Dra
seems an important evidence -- together with the observed one-armed
spiral structure on the rise (section~\ref{structure}) -- in favour
of the MTI model.

\subsection{The uneclipsed component}

\begin{figure}
\centerline{\psfig{figure=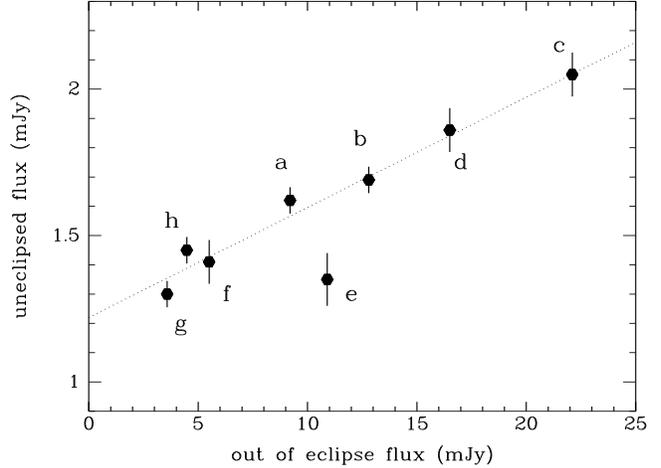,angle=-90,width=9.5cm,rheight=7cm}}
 \caption{ The uneclipsed flux as a function of the out of eclipse 
 brightness level. The labels are the same as in Fig.\,\ref{fig1}. }
\label{fig5}
\end{figure}

The evolution of the uneclipsed component along the outburst can be seen 
in the lower panel of Fig.~\ref{fig1}. The uneclipsed flux is indicated 
by a horizontal tick below each light curve. The horizontal dotted line
marks the value of the uneclipsed flux in the low state.
Our results show that the uneclipsed component is variable and increases 
during outburst in proportion to the overall increase in brightness level.
A plot of the uneclipsed flux as a function of the out of eclipse 
brightness (Fig.~\ref{fig5})
%
shows that there is a clear correlation between these quantities. 
A least-squares linear fit to the data in Fig.~\ref{fig5} (shown as a 
dotted line) indicates that the variable part of the uneclipsed
component corresponds to $\simeq 3$ per cent of the total brightness 
of the system. In the low state, there is a residual uneclipsed flux of
1.3 mJy ($\simeq 36$ per cent of the total $R$-band light at this stage),
which can be understood as the contribution of the red secondary star to
the $R$-band flux. This is the expected $R$-band flux of an M$0\pm2$ 
star at a distance of $\simeq 300\;pc$ (Paper~I).

Assuming that the variable part of the uneclipsed component is zero
in the low state, the variable uneclipsed flux contributes $\simeq 
0.1$~mJy in quiescence, and increases by a factor of $\simeq 8$ at 
outburst maximum.
What is the cause of the variable uneclipsed component?
We discuss two possible explanations.

\subsubsection{A variable disc wind} \label{wind}

Rutten et~al. (1992a) pointed out that when the light curve is 
contaminated by the presence of additional light (e.g., light from the
secondary star or a vertically-extended disc wind) the reconstructed map
shows a spurious ring structure in its back side (i.e., the disc side 
farthest from the secondary star). Since the presence of these
spurious structures is flagged with lower entropy values, 
the correct offset level may be found by comparing a set of maps 
obtained with different offsets and selecting the one with
highest entropy.  Alternatively, the value of the zero-intensity level
can be included in the mapping algorithm as an additional free
parameter (the uneclipsed component) to be fitted along with the 
intensity map in the search for the maximum entropy solution.
Our fitting procedure implements this latter scheme.

Significant (and variable) uneclipsed components were found by
Rutten et~al. (1992b) for the dwarf nova OY~Car in outburst, and by
Baptista et~al. (1996, 1998, 2000) for the nova-like systems UU~Aqr
and UX~UMa. The spectrum of the uneclipsed component in the latter
systems shows pronounced emission lines plus a Balmer jump in emission,
indicating an origin in optically thin gas. In these cases, the 
uneclipsed component was interpreted in terms of emission from a
vertically-extended, largely uneclipsed disc wind. 
The detailed modeling of the C\,{\sc IV} wind line of eclipsing
nova-likes by Schlosman, Vitello \& Mauche (1996) and Knigge \& Drew
(1997) support this scenario. Their results suggest the existence of
a relatively dense ($n_e \sim 4 \times 10^{12}$~cm$^{-3}$) and vertically
extended chromosphere between the disc surface and the fast-moving 
parts of the wind, which could produce significant amounts of optically 
thin emission.

The brightness of EX~Dra during outburst is mostly due to the 
accretion luminosity, $L_{acr} \propto \hbox{\.{M}}_{disc}$. 
If the variable part of the uneclipsed component in EX~Dra is due to
a variable disc wind emission, the linear relation between the 
uneclipsed flux and the total out-of-eclipse brightness indicates that
the wind luminosity is also proportional to the mass accretion rate.

\subsubsection{A flared accretion disc} \label{flared}

Front-back disc brightness asymmetries similar to those found by Rutten 
et~al. (1992a) could also be produced by a flared accretion disc.
In this case, a given surface element in the back side of the disc
appear artificially brighter
than one in the front side due to the different effective areas as
seen by an observer on Earth. Robinson, Wood \& Wade (1999) modeled the 
ultraviolet light curves of the dwarf nova Z~Cha at outburst maximum
assuming a flared disc and found a disc half opening angle of 
$\theta_d= 6\degr$.

Detailed simulations by Wood (1994) show that it is usually impossible 
to distinguish between a flared disc and an uneclipsed component to the 
total light. Eclipse maps obtained with either model may lead to equally 
good fits to the data light curve.

We performed simulations in order to test how a flared disc 
affects an eclipse map obtained with the assumption of a flat disc.
We created steady-state disc brightness distributions for \.{M}$= 
10^{-8}\; M_\odot\,$yr$^{-1}$ at various opening angles and simulated their 
eclipse with the geometry of EX~Dra to construct artificial light curves.
The disc brightness distribution is cut at an outer radius of 
$0.7\;R_{L1}$ and the disc rim is assumed to radiate as a blackbody
with effective temperature equal to the effective temperature of the
disc at the outer radius. The outer disc rim produces a non-negligible
contribution to the total light because it is seen almost face on, which
compensates for the fact that its intrinsic brightness is much lower 
than that of the inner disc due to the lower effective temperature. 
Gaussian noise was added to the light curves
to simulate the signal-to-noise ratio of the real light curves.

The resulting eclipse maps show outer disc rings that become more
pronounced for increasing $\theta_d$. The disc radial brightness 
temperature distribution starts to depart from the $T \propto R^{-3/4}$
law for a steady-state disc when $\theta_d$ is large enough for the inner 
disc regions to be permanently occulted by the disc rim. Outer disc
rings reminiscent of those seen in the eclipse maps $b, c$ and $d$
are produced for $\theta_d = 2-4 \degr$.
The intrinsic front-back asymmetry of a flared disc leads to the 
appearance of a spurious uneclipsed component, the flux of which 
depends on the disc opening angle, $F_{un} \propto {\theta_d}^n$, 
where $n \simeq 0.6-1.0$. 

If the variable part of the uneclipsed component is caused by a 
flaring of the accretion disc of EX Dra, the sensitive dependency
of $F_{un}$ on $\theta_d$ implies a change in disc opening angle
by a factor $\simeq 5-8$ from quiescence to outburst maximum.
Computations of the vertical disc structure (Meyer \& Meyer-Hofmeister
1982; Smak 1992) lead to an estimate of the disc opening angle of
\begin{equation}
\tan \theta_d \simeq 0.038\; \left( 
\frac{\hbox{\.{M}}}{10^{16}\;g\,s^{-1}} \right)^{3/20} \simeq 0.05 \;\; ,
\end{equation}
or $\theta_d \simeq 2-3\degr$ in quiescence (for \.{M}$\simeq 10^{-9}\;
M_\odot\,$yr$^{-1}$). This leads to $\theta_d \simeq 10-24\degr$ at 
outburst maximum which, at an inclination of $i=85\degr$, implies
that at least the inner 1/3 (and up to 2/3, for $\theta_d= 24\degr$)
of the accretion disc would be obscured by the disc rim.
This is hard to reconcile with our observations.
The facts that the radial temperature distribution closely follows
the $T\propto R^{-3/4}$ law for a steady-state disc (even at the 
inner disc regions) and that the regions along the line of sight
to the centre of the disc are much brighter than the outer disc rim
(with a contrast in brightness temperatures of a factor $\simeq 4$)
argue against an obscuration of the inner disc regions at outburst
maximum (map $c$) and indicate that the disc half opening angle at
this stage has to be $\theta_d \simlt 5\degr$.
This is comparable to the value of $\theta_d= 6\degr$ estimated
for Z~Cha at outburst maximum (Robinson et~al. 1999).

Therefore, although there is some evidence of disc flaring in EX~Dra
during outburst, it seems the flaring is not enough to account for the
amplitude of the variation in the uneclipsed component.
However, it may be possible that both disc flaring and variable
uneclipsed wind emission occur during outburst. 
Multi-wavelength eclipse mapping in quiescence and at outburst maximum
could help in better evaluating the importance of each of these effects.
If the variable part of the uneclipsed component is caused by a 
vertically-extended disc wind, the uneclipsed spectrum would show 
a Balmer jump in emission plus strong emission lines, the strength
of which should increase in response to the overall brightness
(i.e., disc mass accretion) increase during outburst.
In the case of a flared disc the spurious uneclipsed spectrum should
reflect the difference between the disc spectrum of the back (deeper
atmospheric layers seen at lower effective inclinations) and the
front (upper atmospheric layers seen at grazing incidence) sides and
should mainly consist of continuum emission filled with absorption
lines.

\subsection{The inner disc on the rise}

The left-hand panels of Fig.~\ref{fig3} show that the brightness of 
the inner disc regions of EX~Dra {\em decrease} during the rise 
(from maps $h$ to $a$ and to $b$).
This is in contrast with the results from OY~Car, which displays 
a continuous increase in the brightness of the inner disc regions
from quiescence to outburst maximum (Rutten et~al. 1992b).
This distinct behaviour of EX~Dra is possibly a consequence 
of its larger mass transfer rate.
We discuss three possible explanations for the observed effect.

\subsubsection {An intrinsic effect}

In the DI model, if the quiescent inner disc remains in a high 
viscosity, steady-state (see section~\ref{quiescent}), the outburst
starts at the interface between the inner hot and the outer cool disc 
(e.g., Lin et~al. 1985).
In this case, the outward-moving heating front drives a mass outflow 
from the interface region possibly reducing the gas supply that 
feeds the inner, steady-state disc. This could lead to a temporary 
reduction in mass accretion in the inner disc regions causing a
decrease in the brightness of these regions. However, there is no
support for this behaviour in numerical simulations of outbursting 
accretion discs so far.

\subsubsection {An optical depth effect}

If the inner disc is optically thin in quiescence (see 
section~\ref{quiescent}), both hemispheres of the central source 
(e.g., the white dwarf and the boundary layer) are visible. 
It is possible that the inner disc becomes optically thick on the rise, 
covering the lower hemisphere of the central source and leading to a 
reduction of a factor $\simeq 2$ on its flux. Of course, this explanation
fails if the `central source' is, in fact, an inner disc always in an 
optically thick steady-state, as suggested in section~\ref{quiescent}.
The development of an optically-thick wind from the inner disc regions 
during the rise could produce a similar effect, although this would 
lead to the question of why the effect is more pronounced during the
rise and not at outburst maximum, when the wind is probably stronger.

\subsubsection {An obscuration effect}
It may be that the accretion disc thickens significantly during the rise
as a consequence of the mass outflow produced by the outward-moving
heating front (e.g., Lin et~al. 1985; Menou et~al. 1999). 
In this case, the apparent reduction in brightness of the inner disc may 
be caused by obscuration of the inner disc regions by the thick disc rim. 
An interesting consequence of this alternative is that the observed 
outburst maximum (map $c$) would not necessarily reflect the time of 
maximum disc brightness, but the time when the disc rim becomes thinner
(possibly as the disc reaches a hot, steady-state) and the inner regions 
start being visible again.

While this is a plausible explanation, it is hard to reconcile with 
the results of the simulations with flared discs of section~\ref{flared}
because it requires that the maximum of the (spurious) uneclipsed light
occurs during the rise and not at outburst maximum, as observed
(Fig.~\ref{fig1}).

\section{Conclusions} \label{fim}

We have used eclipse mapping techniques to study the structure and
the time evolution of the accretion disc of EX~Dra throughout its 
outburst cycle. 
The maps show the formation of a one-armed spiral structure on the
rise to outburst. The evolution of the radial intensity distribution
of the maps throughout the cycle shows that the disc expands during the 
rise. It also suggests the presence of an inward and an outward-moving
heating wave during the rise and an inward-moving cooling wave in the
decline.
The disc becomes progressively fainter during the decline until only a
small bright region around the white dwarf is left at minimum light. 

We were able to derive estimates of the velocities of the heating
and cooling waves from the maps. These were found to be in
reasonable agreement with predictions from numerical simulations
of transition fronts in accretion discs.
We observe a systematic deceleration of both the heating and the
cooling waves in support of the DI model.

Our analysis of the brightness temperature profiles indicates that most 
of the disc appears to be in steady-state during quiescence and at outburst
maximum, but not during the intermediate stages. 
As a general trend, the mass accretion rate in the outer regions is 
larger than in the inner disc on the rising branch, while the opposite 
holds during the decline branch.
We infer a mass accretion rate of about $10^{-8}\,M_\odot$\,yr$^{-1}$ 
at outburst maximum and $10^{-9.1}\, M_\odot\,$yr$^{-1}$ in quiescence.
The analysis of $V$ and $R$-band brightness temperature distributions
in quiescence suggests that the viscosity parameter is high at this
stage, $\alpha_{cool}\simgt 0.25$, which favours the MTI model.

We have found an uneclipsed source of light in all the maps.
The uneclipsed light has a steady component arising from emission from
the red dwarf and a variable component which varies linearly with the 
overall out-of-eclipse flux and corresponds to about 3 per cent of the 
total brightness of the system.
Although disc flaring is likely in EX~Dra during outburst, it
is not enough to account for the amplitude of the variation of
the uneclipsed source.
Thus, we interpret the variable component as mainly due to emission 
arising from a disc wind.

The brightness of the inner disc regions decrease during the rise.
This may be an intrinsic effect (temporary reduction of mass accretion
in the inner disc), an optical depth effect (the inner disc becomes
optically thick and partially blocks the light from the central source),
or an obscuration effect (a thick disc rim hidding the inner disc from
view).

\section*{Acknowledgments}

We thank an anonymous referee for useful comments that improved 
the presentation of the paper.
In this research we have used, and acknowledge with thanks, data from
the AAVSO International Database and the VSNET that are based on
observations collected by variable star observers worldwide. 
This work was partially supported by the PRONEX/Brazil program through
the research grant FAURGS/FINEP 7697.1003.00. RB acknowledges financial 
support from CNPq/Brazil through grant no. 300\,354/96-7.
MSC acknowledges financial support from a PPARC post-doctoral grant 
during part of this work.

\appendix
\section{Reconstructions from light curves of incomplete phase coverage}
\label{apendice}

In an eclipse mapping reconstruction, each surface element in the 
accretion disc is uniquely labeled by its ingress and egress phases.
The brightness of a given surface element is derived from the combined
information given by the changes in flux caused by its occultation 
(at ingress) and reappearance (at egress) from behind the secondary star.
In the case of an eclipse light curve with incomplete phase coverage,
there are regions in the disc for which only one of these pieces of 
information is available.

We performed simulations in order to assess the reliability of eclipse 
mapping reconstructions obtained from light curves of incomplete phase 
coverage such as the mid-decline light curve $e$.

\begin{figure*}
\centerline{\psfig{figure=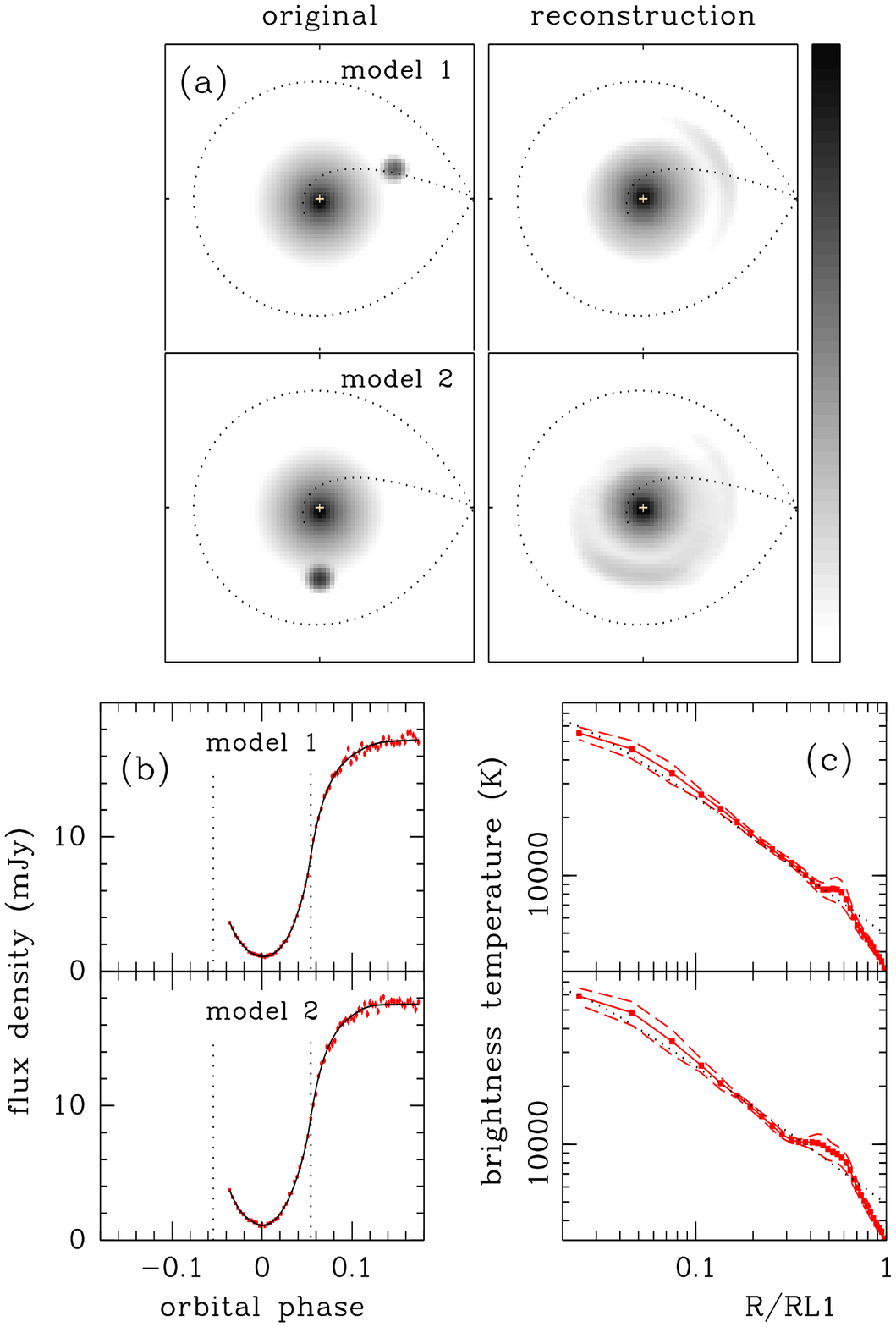,width=16cm,rheight=20cm}}
 \caption{ Simulations of eclipse mapping reconstructions for the case
 of incomplete phase coverage.
 (a) The left-hand panels show two artificial brightness distributions in
 a logarithmic greyscale. These are steady-state disc models with
 \.{M}$=10^{-8}\;M_\odot$\,yr$^{-1}$ and Gaussian bright spots
 added at different azimuths. The corresponding reconstructions are shown 
 in the right-hand panels. 
 (b) The synthetic light curves from the brightness distributions 1 and 2 
 (dots) and the eclipse mapping models (solid lines). Vertical 
 dotted lines mark the ingress and egress phases of the central source. 
 (c) The resulting radial brightness temperature distributions. 
 A steady-state disc model for a mass accretion rate of 
 \.{M}$=10^{-8}\;M_\odot$\,yr$^{-1}$ is plotted as a dotted line for
 comparison. The notation is the same as in Fig.~\ref{fig4}. }
\label{fig6}
\end{figure*}

For this purpose, we constructed two artificial brightness distributions 
using a steady-state disc model with \.{M}$=10^{-8}\;M_\odot$\,yr$^{-1}$, 
$M_1= 0.75\; M_\odot$, $R_1= 0.011\;R_\odot$, and $R_{L1}= 0.85\;R_\odot$. 
These parameters reproduce those of the accretion disc of EX~Dra at 
outburst maximum (see section~\ref{trad}). The artificial brightness
distributions are shown in Fig.~\ref{fig6} in a logarithmic greyscale.
For one of them (model 1) we added a Gaussian spot at the expected 
position of the bright spot in EX~Dra, along the gas stream trajectory. 
For the other (model 2), we added a similar Gaussian spot in the leading 
side of the disc (the lower hemisphere of the maps shown in Fig.~\ref{fig6}). 
We adopted the geometry of EX~Dra ($q=0.72$ and $i=85\degr$) and 
constructed light curves with the same signal-to-noise ratio and 
(incomplete) orbital phases of the light curve $e$.
The artificial light curves were fitted with the eclipse mapping code 
used in section~\ref{mem}. The results are shown in Fig.~\ref{fig6}.
%

For the set of orbital phases of light curve $e$, the trailing side of 
the disc (the upper hemisphere of the eclipse maps in Fig.~\ref{fig6}) 
is mapped by the moving shadow of the secondary star both during ingress
and egress, whereas most of the leading side of the disc is only mapped
by the secondary star at egress phases.

Despite the incomplete coverage of the eclipse, the reconstruction of
model 1 shows an asymmetric structure at the correct position of the 
bright spot. The structure is elongated in azimuth because of the
intrinsic azimuthal smearing effect of the eclipse mapping method.
There is good agreement between the original and the reconstructed
radial brightness temperature distributions.
Similarly good results are obtained for model~2, despite the fact that
the spot in this case is located in the disc region for which there is 
limited information in the shape of the light curve. The azimuthal 
smearing effect is more pronounced than for model~1 because, in this 
case, the brightness distribution is less constrained by the data.
Nevertheless, the radial position of the spot and the radial temperature
distribution are well recovered.
Similar results were obtained for light curves of comparable incomplete 
phase coverage at egress.

We performed additional simulations by adding artificial offsets to the 
light curves in order to test the ability of the eclipse mapping method to
recover an uneclipsed component for the case of light curves of incomplete 
phase coverage. The results indicate that the ability to recover the
uneclipsed component is as good as in the case of a light curve with 
complete phase coverage.

Therefore, we conclude that the results inferred from the light curve 
$e$ are as reliable as those from the other light curves in our study.

\bsp

\end{document}